# Reliability of Magneto-elastic Switching of Non-ideal Nanomagnets with Defects: A Case Study for the Viability of Straintronic Logic and Memory

David Winters[1], Md Ahsanul Abeed[1], Sourav Sahoo[2], Anjan Barman[2] and Supriyo Bandyopadhyay[1*]

[1]Dept. of Electrical and Computer Engineering, Virginia Commonwealth University, Richmond, VA 23284, USA

[2]Dept. of Condensed Matter Physics and Materials Science, S. N. Bose National Center for Basic Sciences, Kolkata 700116, India

## Abstract

Magneto-elastic (straintronic) switching of bistable magnetostrictive nanomagnets is an extremely energy-efficient switching methodology for (magnetic) binary switches that has recently attracted widespread attention because of its potential application in ultra-low-power digital computing hardware. Unfortunately, this modality of switching is also very error prone at room temperature. Theoretical studies of switching error probability of magneto-elastic switches have predicted probabilities ranging from $10^{-8} - 10^{-3}$ at room temperature for *ideal*, *defect-free* nanomagnets, but experiments with real nanomagnets show a much higher probability that exceeds 0.1 in some cases. The obvious spoilers that can cause this large difference are defects and non-idealities. Here, we have theoretically studied the effect of common defects (that occur during fabrication) on magneto-elastic switching probability in the presence of room-temperature thermal noise. Surprisingly, we found that even small defects increase the switching error probabilities by orders of magnitude. There is usually a critical stress that leads to the lowest error probability and its value increases enormously in the presence of defects. All this could limit or preclude the application of magneto-elastic (straintronic) binary switches in either Boolean logic or memory, despite their excellent energy-efficiency, and restrict them to non-Boolean (e.g. neuromorphic, stochastic) computing applications. We also studied the difference between magneto-elastic switching with a stress pulse of constant amplitude and sinusoidal time-varying amplitude (e.g. due to a surface acoustic wave) and found that the latter method is more reliable and generates lower switching error probabilities in most cases, provided the time variation is reasonably slow.

[*]Corresponding author: email: sbandy@vcu.edu

I. **Introduction**

There is a strong belief in the device physics community that the celebrated Moore's law of electronics is nearing its end and alternate device paradigms must replace the famed "transistor" in order to continue downscaling of computing and signal processing devices. Devices that exploit the spin degree of freedom, as opposed to the charge degree of freedom, (e.g. bistable nanomagnets) are a potential candidate. The energy that is dissipated in switching a bistable nanomagnet depends on the switching methodology and the one that has been found to be remarkably energy-efficient is magneto-elastic (or "straintronic") switching [1]. Straintronic switches have been advocated as a serious contender to replace transistors [2].

The energy dissipated in a magneto-elastic switching event, that takes place over ~1 ns, can be less than 1 aJ at room temperature [1], which eclipses most other switching methodologies. The archetypal magneto-elastic switch is composed of a magnetostrictive nanomagnet delineated on a piezoelectric substrate, which is strained with a small voltage of few mV to generate stress in the nanomagnet [3-8]. This stress makes the magnetization vector rotate from one stable direction to the other and makes the nanomagnet "switch" [9-11]. Unfortunately, this switching mechanism is also error prone [12-16]. At room temperature, the switching failure probability in ideal and pristine defect-free nanomagnets has been calculated to be as high as $10^{-3}$ and as low as $10^{-8}$ [12-19]. This is too high for mainstream Boolean logic where the switching error probability should be less than $10^{-15}$ [20]. Memory is more forgiving than logic, but the write error probability in memory chips, determined by the switching error probability, should not exceed $10^{-9}$ [21], which makes even memory applications questionable.

Experiments tell an even more foreboding tale. The error probability (failure to switch) in fabricated nanomagnets was found to be as high as 0.75 in some cases [22-24]. Obviously, this would preclude any application in logic or memory, and restrict magneto-elastic switches to non-Boolean computing, e.g. Bayesian inference engines [25, 26], image processing [27, 28], ternary content addressable memory [29], restricted Boltzmann machines [30]) and sub-wavelength antennas [31-33] to name a few. There, a high error rate may be tolerable and the excellent energy efficiency is a welcome boon.

The purpose of this paper is twofold: to determine what causes the large difference between theoretical estimates of the error probability and experimental observations, and to determine if the cause can be ameliorated. If not, then the error rates will be too large for both logic and memory applications (mainstream applications) and restrict magneto-elastic switches to specific non-Boolean (niche) applications. Clearly, the difference between ideal nanomagnets and real nanomagnets is that the latter have structural defects acquired during the fabrication process. Recently, we theoretically simulated the magneto-elastic switching of a *non-ideal* nanomagnet containing a small "hole" in the center in the presence of thermal noise and found that the small defect can vastly increase the switching error probability [34]. In this paper, we have extended that study to other types of common defects, including extended defects that are inherently different in influence than the localized point defect considered in [34], and also considered the effect of time-varying stress as opposed to static stress in various defective nanomagnets. We found that in all cases, defects increase the switching error probability by *orders of magnitude* and this could at least partially explain why the error rates found in experiments vastly exceed those estimated from theoretical simulations of ideal specimens. This is a discouraging result and implies that unless pristine nanomagnets can be fabricated routinely, magneto-elastic switches may not be suitable for Boolean logic and memory. This would temper some of the enthusiasm about magneto-elastic switches.

## II.     Nanomagnets with structural defects

Fig. 1 shows atomic force micrographs of some Co nanomagnets fabricated in our lab. They are delineated on a piezoelectric substrate (with rms surface roughness 1-3 nm) for magneto-elastic switching. These nanomagnets were fabricated by patterning a PMMA electron beam resist (spun on to the piezoelectric substrate) with e-beam lithography. The resist was developed and cobalt was evaporated within opened windows using electron beam evaporation, followed by lift off, to produce the nanomagnets. Normally, such nanomagnets cannot be fabricated with ion-milling since the damage to the piezoelectric substrate caused by that process would be intolerable. Ion milling is routinely used to fabricate magneto-tunneling junctions switched with spin-transfer torque, but there the substrate is non-piezoelectric and hence not bound by the same constraints.

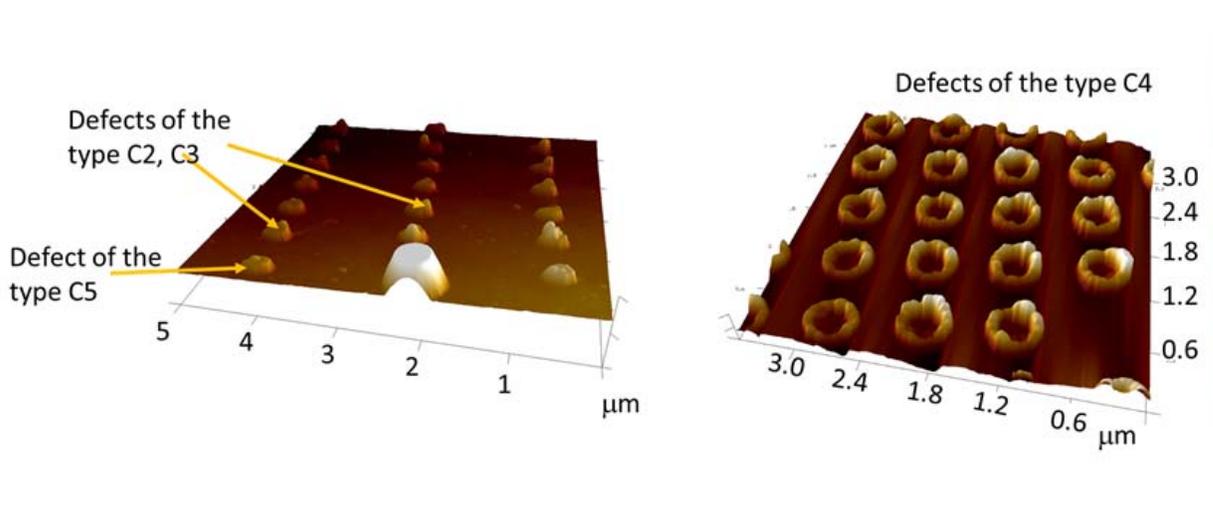

Fig. 1: Atomic force micrographs of nanomagnets showing the various types of thickness variations across the surface that can result from the fabrication process.

In all cases, we observed thickness variations across the surface of the nanomagnets. These variations are not specific to a given fabrication run, but show up in every run, although there are obviously slight variations between different runs. Some of them may be caused by the large surface roughness in piezoelectric substrates, which is typically ~1 nm. This is much worse than the surface roughness (< 0.3 nm) found in silicon substrates used to fabricate spin transfer torque memory. Unfortunately, magneto-elastic switches require a piezoelectric film or substrate which has a relatively large surface roughness. Thus, these defects might be unavoidable even under the most stringent fabrication control. We have classified the observed defects into six different classes, each one of which is approximated in the manner of Fig. 2.

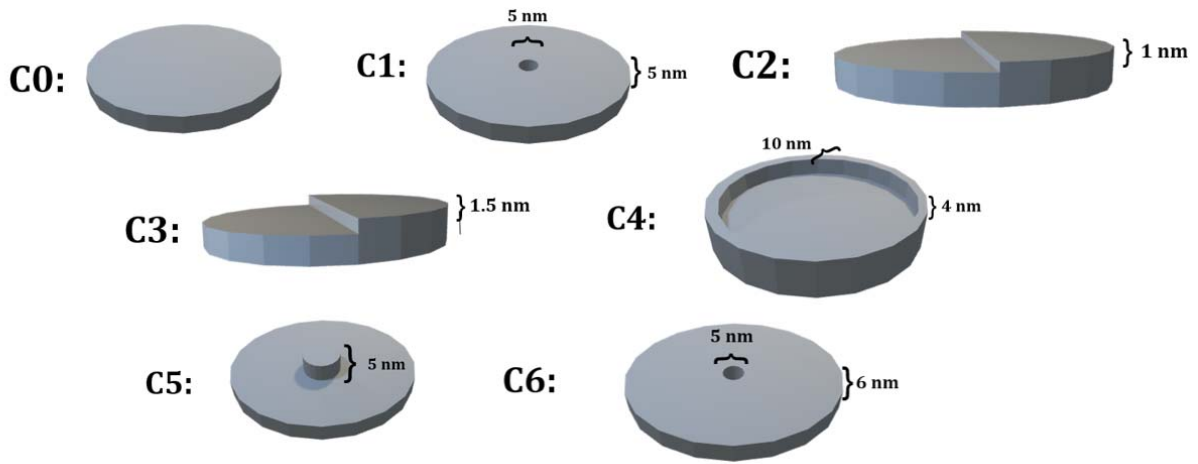

Fig. 2: The different types of thickness variations observed in Fig. 1 are approximated by six different configurations: C0 (no defect, an elliptical disk of major axis 100 nm, minor axis 90 nm and thickness 6 nm), C1 (a shallow hole 5 nm in diameter and 5 nm deep at the center, not observed in Fig. 1, but still commonplace), C2 and C3 (one half of the nanomagnet thicker than the other by 1 nm and 1.5 nm respectively), C4 (an annulus 10 nm thick and 4 nm high at the periphery; we kept the height and thickness uniform for ease of simulation. Later we show that this defect is the worst and increases the switching error probability dramatically. Introducing randomness in the height and thickness of the annulus will, if anything, exacerbate the error), C5 (a raised cylindrical region 5nm in diameter and 5 nm high), and C6 (a through hole 5 nm in diameter, not observed in Fig. 1).

### III. Simulation of magneto-elastic switching of defect-free and defective nanomagnets

We simulate the switching dynamics of the seven different nanomagnets depicted in Figure 2 in the presence of room-temperature thermal noise under the switching scenario shown in Fig. 3. These nanomagnets contain the type of defects that are experimentally observed and shown in Fig. 1. All seven nanomagnets are elliptical disks of major axis 100 nm, minor axis 90 nm and nominal thickness 6 nm. The two stable magnetization directions are along the major axis (easy axis) pointing in mutually opposite directions. The nanomagnets are amorphous or polycrystalline (since they are delineated by metal evaporation) and therefore we do not consider crystalline defects. We assume that the magnetization is initially pointing in one of these directions and a magnetic field of 3 mT is applied in the opposite direction to make the magnetization flip (switch), as shown in the left panel of Fig. 3. This magnetic field need not be a real field, but could be the dipolar coupling field of a nearby nanomagnet (not shown).

Simulations of the temporal evolution of the magnetization vector in defective and defect-free nanomagnets under the applied bias magnetic field and stress were performed using the micromagnetic program Mumax3. Mumax3 solves the Landau-Lifshitz-Gilbert equation (LLG) for the switching dynamics taking into account the effective magnetic fields acting on the magnetization due to shape anisotropy $\left(\vec{H}_{shape}\right)$, the bias magnetic field $\left(\vec{H}_{bias}\right)$, the exchange interaction field $\left(\vec{H}_{ex}\right)$, the strain $\left(\vec{H}_{strain}\right)$, and random thermal noise $\left(\vec{H}_{th}\right)$. Descriptions of the simulation procedure and further details can be found in ref. [34], and hence not repeated here. The total magnetic field is given by

$$\vec{H} = \vec{H}_{shape} + \vec{H}_{bias} + \vec{H}_{ex} + \vec{H}_{strain} + \vec{H}_{th} \qquad (1)$$

We ignore magneto-crystalline anisotropy since the nanomagnets are amorphous. The mesh size used is 2 nm, which is smaller than the magnetic coherence length (the simulation results do not change if we make the mesh smaller) and the time step used in the simulation is 0.1 ps.

We turn on the magnetic field at time $t = 0$, allow the micromagnetic distribution within the nanomagnet to reach steady-state and then turn on uniaxial stress along the major axis of the ellipse, either of a fixed amplitude or sinusoidal time-varying amplitude, for a given duration. In the case of fixed amplitude stress, the duration is long enough to ensure that the micromagnetic distributions have reached steady state under the applied stress. In the case of sinusoidal stress, the duration is one period of the sinusoid. We simulate the temporal evolution of the magnetization vector using the micromagnetic simulator MuMax3 and continue the simulation after stress withdrawal, until steady state is reached. The final steady state orientation of the magnetization vector at the end of the simulation ends up being either nearly parallel or nearly antiparallel to the magnetic field. That tells us whether switching was successful (magnetization pointing close to the direction of the magnetic field) or failed (magnetization pointing close to the opposite direction). This allows us to determine the switching error probability by running the simulation a number of times and determining what fraction of the trials ended in failure. That fraction is the error probability. We study the error probability for various types of defects as a function of stress amplitude for both pulsed and sinusoidal stress waveforms. These studies reveal interesting dependences of the error probability on various parameters, but ultimately show that the error probabilities may be too high for both logic and memory.

IV. **Switching error probability**

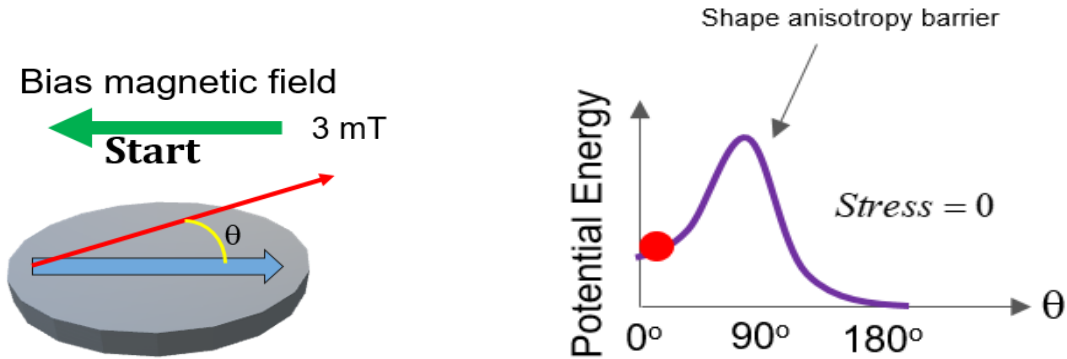

Fig. 3: (Left) A defect-free nanomagnet in the form of an elliptical disk. The magnetization initially points to the right and a 3 mT magnetic field is applied to make it flip to the left. (Right) The potential energy profile of the nanomagnet in the presence of a magnetic field plotted as a function of the in-plane angle θ that the magnetization subtends with the initial direction. There is a local minimum at θ = 0° and a global minimum at θ = 180° due to the bias magnetic field. The magnetization is initially at the local minimum shown by the ball. The energy barrier is due to shape anisotropy. This energy barrier prevents the magnetization from immediately flipping to the left and reaching the global minimum. Stress will erode or invert this barrier to allow the magnetization to reach the global minimum and flip in the direction of the applied magnetic field.

In the switching scenario that we have studied, the bias magnetic field alone cannot flip the magnetization from right to left in the defect-free nanomagnet in Fig. 3 since the shape anisotropy energy barrier within the nanomagnet is designed to be too high for the magnetic field to overcome. This is shown in the right panel of the cartoon in Figure 3. The nanomagnet is stuck in the local potential minimum corresponding to the original magnetization direction (θ = 0°) and cannot get to the global minimum at θ = 180° to complete the switching because of the intervening potential barrier. Uniaxial stress applied along the major axis of the elliptical nanomagnet lowers or inverts the barrier as shown in the cartoon of Fig. 4. Once stress inverts the barrier, the magnetization migrates to the new energy minimum that forms under stress and points along the minor axis of the nanomagnet, as shown in the middle panel of Figure 4. After stress is removed, the original potential profile is restored, as shown in the right panel of Figure 4. At that point, the magnetization could swing either to the left or to the right orientation (with a higher probability of swinging to the left because of the magnetic field which makes the energy minimum at θ = 180º lower than that at θ = 0º), but the probability of swinging to the right is non-zero. The latter probability (probability of failure to switch) decreases as we make the energy difference between the local and global minima larger by making the magnetic field stronger, but the failure probability never vanishes. It is this failure probability (or error probability) that we calculate for the six different types of defects enumerated in Fig. 2, as well as the defect-free nanomagnet.

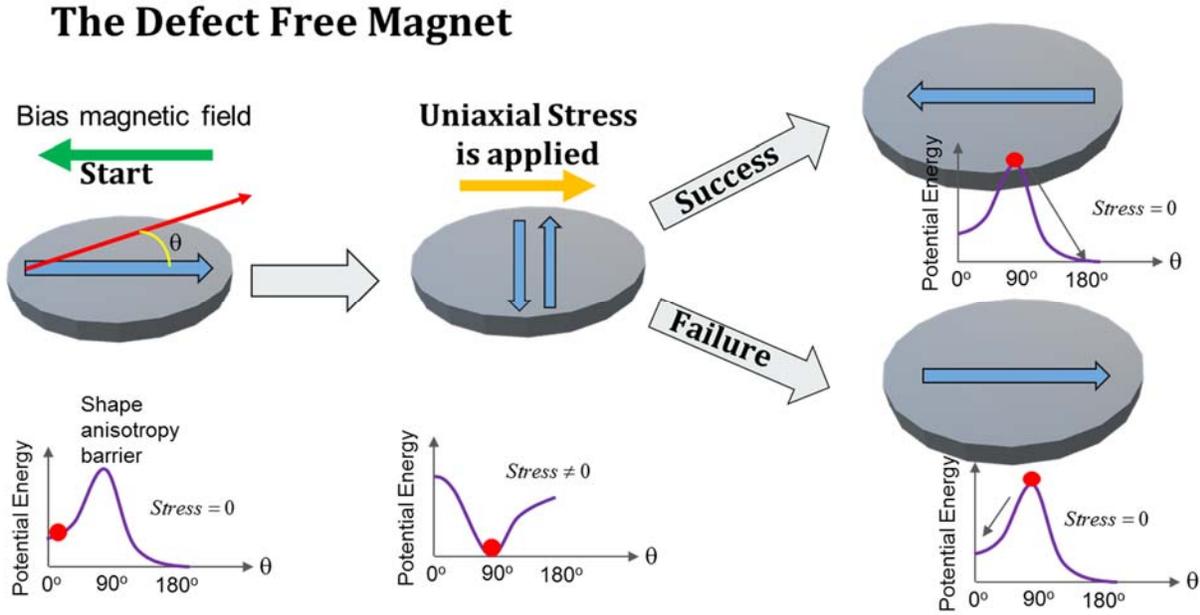

Fig. 4: Left panel: The initial state with the magnetization pointing to the right along the major (easy) axis of magnetization and a magnetic field point to the left. The magnetic field cannot flip the magnetization taking the system from the local to the global energy minimum because of the intervening shape anisotropy energy barrier. Middle panel: Sufficiently strong uniaxial stress applied along the major axis inverts the potential barrier and makes the magnetization point along the minor axis. Right panel: Upon stress release, the magnetization can flip either to the stable state on the left or to the stable state on the right, with the probability of the latter being lower owing to the magnetic field which makes the global minimum lower in energy than the local minimum.

V.  **Critical stress**

The switching error probability usually has a *non-monotonic* dependence on the magnitude of stress. The cause can be easily understood for the defect-free case and is explained by the cartoon in Fig. 5. If we do not apply sufficient stress to erode the barrier (*under-stressed*), then the likelihood of the magnetic field overcoming the barrier to switch the magnetization will be low. After stress is released, the magnetization will very likely remain in its original direction, resulting in a large switching error probability as shown in the left panel of Fig. 5. If we *over-stress* to not only erode, but actually invert, the barrier, then the magnetization will rotate to align along the hard axis ($\theta = 90^0$). Upon stress removal, the magnetization will temporarily find itself at the unstable energy maximum at $\theta = 90^0$. From there, it will have a higher probability of switching correctly in the direction of the bias magnetic field because the latter makes the minimum at $\theta = 180°$ lower in energy than that at $\theta = 0^0$. However, the failure probability (ending up at $\theta = 0^0$ instead of at $\theta = 180°$) will still be non-zero, as shown in the middle panel of Fig. 5. On the other hand, if we apply just the right amount of stress (*critical stress*) that erodes *but does not invert* the potential barrier, then the magnetization state will smoothly transition from the initial local minimum at $\theta = 0°$ to the global minimum at $\theta = 180°$. Thereafter, when stress is released, the magnetization will find itself pointing in the direction of the magnetic field with the highest likelihood as shown in the right panel of Figure 5. Thus, the critical stress will yield the *minimum* error probability.

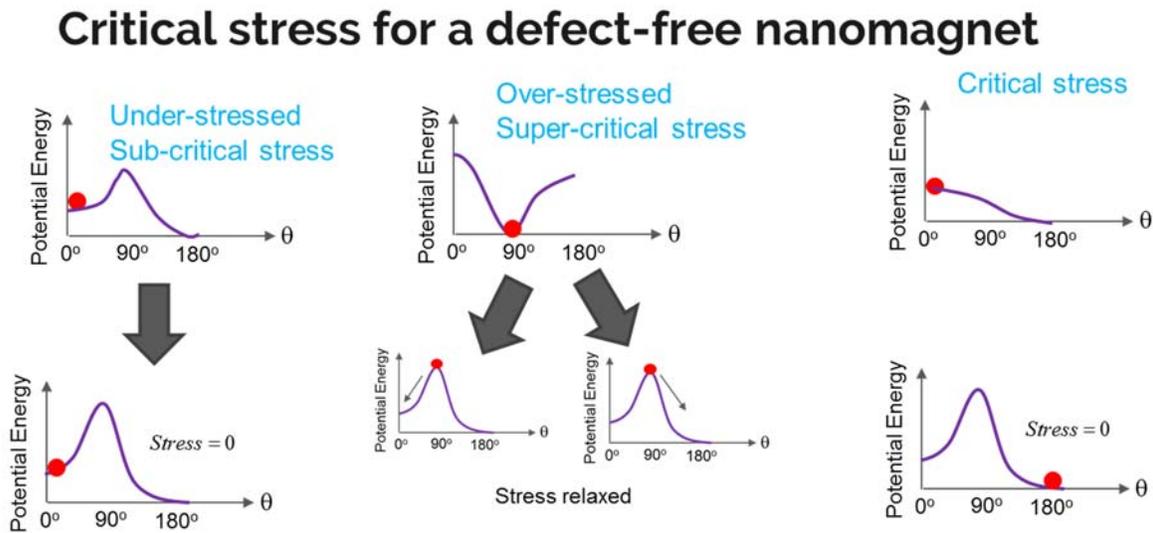

Fig. 5: Left panel: Sub-critical stress does not erode the barrier and the switching error probability remains high after stress is removed. Middle panel: Super-critical stress inverts the barrier and upon stress removal, the magnetization has a higher probability of switching successfully than failing to switch, but the failure probability is not insignificant. Right panel: Critical stress just erodes the potential barrier without inverting it and the smoothly transitions to the global energy minimum, so that after stress release, the magnetization would have flipped in the direction of the magnetic field with maximum probability.

## VI. Micromagnetic simulations of magneto-dynamics in defective and defect-free nanomagnets to find switching error probabilities as a function of stress.

In the MuMax3 simulation, the magnetic field is turned on at time $t = 0$ and the micromagnetic distribution within a nanomagnet (either defective or defect-free) is allowed to reach steady state. In all cases, steady state is reached within 1 ns. Once that happens, uniaxial stress (compressive for a material with positive magnetostriction and tensile for a material with negative magnetostriction) is applied along the major axis of the elliptical nanomagnet for a duration of 1.5 ns, and the temporal dynamics of the magnetization is monitored. This duration of 1.5 ns is long enough for the micromagnetic distributions to reach steady state under stress. The stress is withdrawn after 1.5 ns and the magnetization is once again allowed to reach steady state, which takes another 1.5 ns or less. When the final steady-state is reached, the normalized component of the magnetization vector along the major axis will almost always be either within the interval [-1, -0.85] or within the interval [0.85, 1]. In the former case, we conclude that the magnetization has succeeded in switching from right to left under the action of the bias magnetic field and in all other cases, we conclude that the magnetization has failed to switch.

We simulate 1000 switching trajectories for each case as described above. Each trajectory is slightly different from all others because of the random thermal noise. The fraction of the trajectories that end up in failure is the *error probability*. We simulate the magnetization dynamics in all seven nanomagnets depicted in Fig. 2 for two different materials – Terfenol-D and cobalt. The material parameters used are listed in Table I [35-38]. The energy barriers in the defect-free nanomagnets are 32 kT (Terfenol-D) and 60.5 kT (Cobalt) at room temperature.

There is one rare but curious outcome in our simulations that is different from the rest. In the cases of defects C2 and C3 in Terfenol-D, we found that the magnetization vector occasionally finds a stable orientation along the minor axis and remains pinned there. This can never happen in the case of the defect-free nanomagnet since the minor axis will be the hard axis (energy maximum) and thermal perturbation will inevitably dislodge the magnetization from there, but in the case of *defective* nanomagnets, shallow energy minima may be created for other orientations (including along the minor axis) and the magnetization may be trapped there. Thermal perturbations may not always be able to dislodge it from these metastable states. We stress that in defective nanomagnets, the potential energy profile may not look anything like those in Figures 4 or 5 and may have one or more local minima that pin the magnetization and increase the switching error probability. In most cases however, the steady-state magnetization will assume an orientation close to the major axis so that the component along the major axis will be either within the interval [-1, -0.85] or the interval [0.85, 1].

**Table I: Material parameters**

|  | **Terfenol-D** | **Cobalt** |
|---|---|---|
| Saturation magnetization (A/m) | $8 \times 10^5$ | $1.1 \times 10^6$ |
| Gilbert damping constant | 0.1 | 0.01 |
| Exchange constant (J/m) | $9 \times 10^{-12}$ | $3 \times 10^{-11}$ |
| Saturation magnetostriction (ppm) | +600 | -35 |

In the cases of defects C2 and C3 in cobalt, the energy barrier in the nanomagnet is so low that the magnetic field alone can overcome it and make the magnetization flip, without the need for any stress. Hence these two cases are not simulated in the case of cobalt. However, this is not the case with Terfenol-D.

We also studied the case of sinusoidal time-varying uniaxial stress applied along the major axis of the ellipse since it has emerged as a common method of magneto-elastic switching [39-50]. We apply the bias magnetic field, allow the micromagnetic distributions to reach steady-state, then run the simulations for one full cycle of the sinusoid, and finally remove the stress and allow the magnetization vector to reach steady state. Once again, the steady-state magnetization will assume an orientation close to the major axis so that the component along the major axis will be either within the interval [-1, -0.85] or the interval [0.85, 1].

We considered sinusoidal stresses of frequencies 200 MHz and 1 GHz, and various amplitudes. Frequencies lower than 200 MHz are computationally prohibitive for us since in order to cover one cycle, we have to run the simulations for 5 ns for each of the 1000 trajectories for each defect, which consumes significant computer resources. For the sinusoidal stress, we restrict ourselves only to cobalt (not Terfenol-D) and do not consider defects C2 and C3 since no stress is required to switch the magnetizations of nanomagnets containing these defects, as mentioned before.

### VII.    Results and Discussion

In Figure 6, we plot the switching error probability (i.e. failure of a nanomagnet to switch its magnetization from right to left under the action of the left pointing bias magnetic field) as a function of abrupt uniaxial stress applied along the major axis of the nanomagnet for a fixed duration of 1.5 ns (a longer duration will not be fruitful since the micromagnetic distributions reach steady state under stress in less than 1.5 ns). The material is *Terfenol-D* and the plots are for the defect-free (C0) and defective (C1-C6) cases. The maximum applied stress amplitude is 50 MPa. The Young's modulus of Terfenol-D is about 80

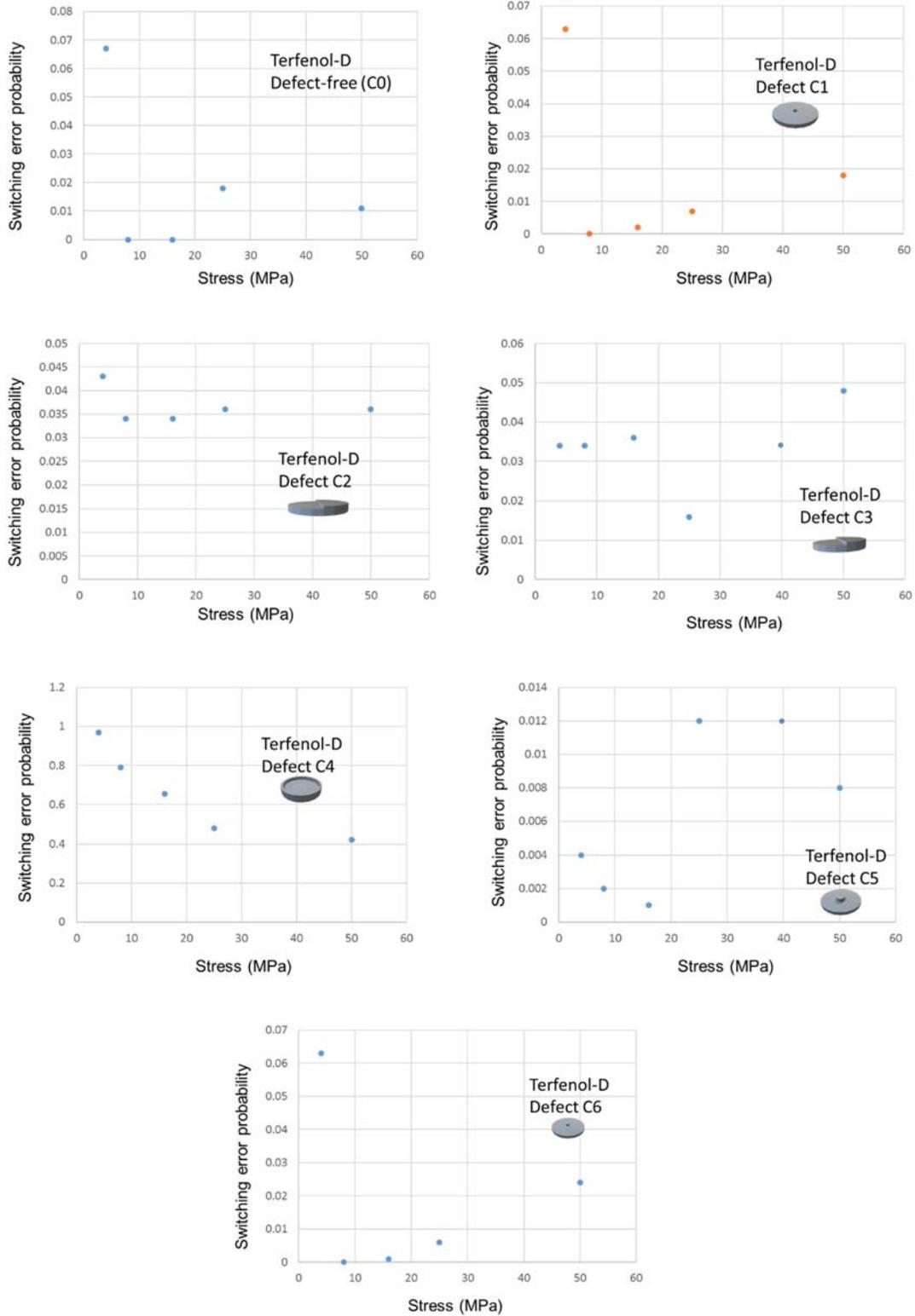

Fig. 6: Switching error probability at room temperature as a function of applied uniaxial stress for a Terfenol-D elliptical nanomagnet of nominal major axis 100 nm, minor axis 90 nm and thickness 6 nm. The plots are for the defect-free (C0) and defective nanomagnets with six different defects (C1-C6).

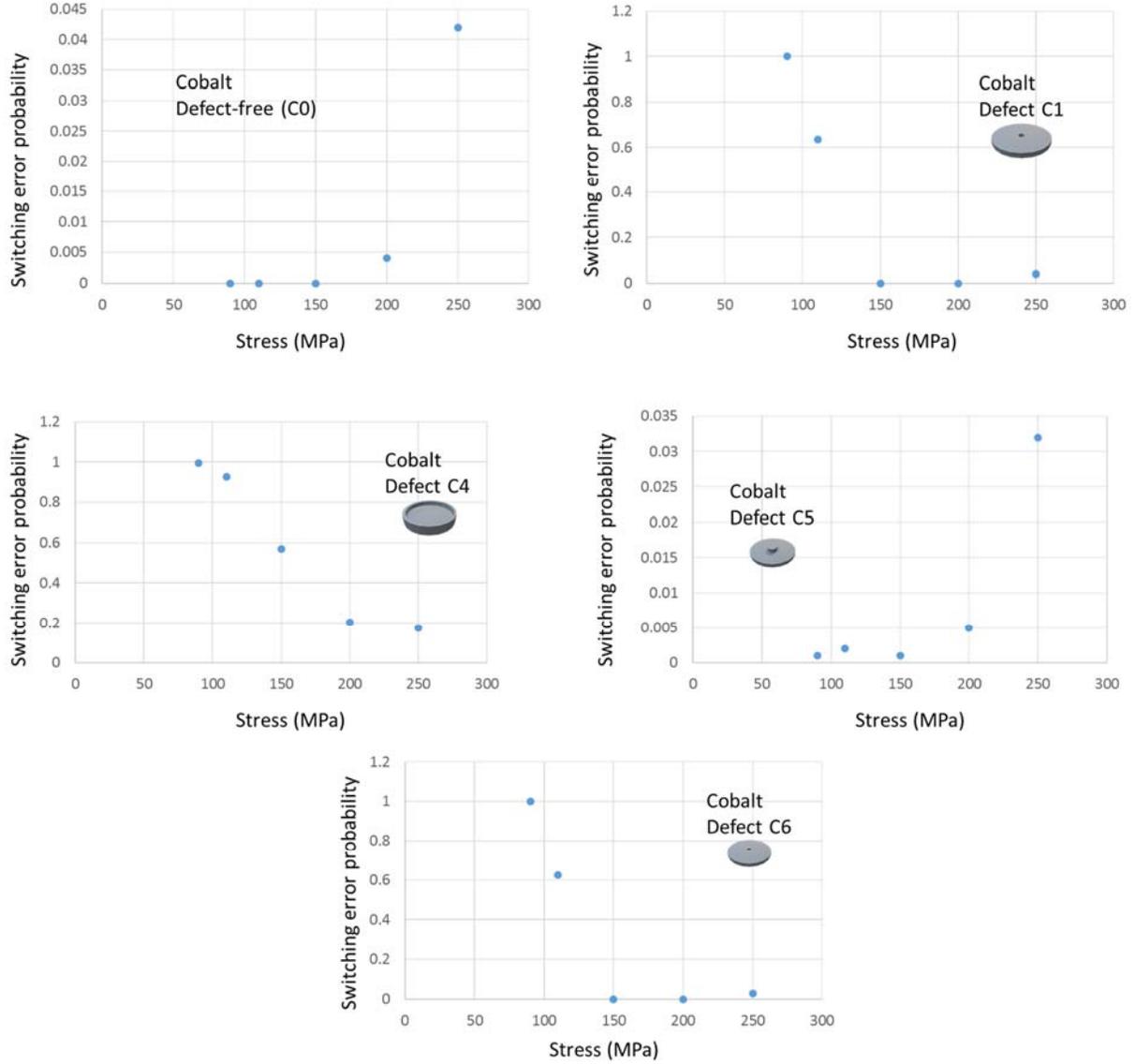

Fig. 7: Switching error probability at room temperature as a function of applied uniaxial stress for a cobalt elliptical nanomagnet of nominal major axis 100 nm, minor axis 90 nm and thickness 6 nm. The plots are for the defect-free (C0) and defective nanomagnets with defects C1, C4, C5 and C6. For C0 and C5, the error probabilities at 50 MPa stress are 0.997 and 0.998 respectively.

GPa [38] and hence a 50 MPa stress would produce a strain of 625 ppm which is about the maximum a Terfenol-D nanomagnet can tolerate before stress relaxes via the formation of cracks and dislocations. We can calculate an effective magnetic field $H_{eff}$ for switching that such a stress would generate from the relation $\mu_0 M_s H_{eff} = (3/2)\lambda\sigma$, where $\mu_0$ is the magnetic permeability of free space, $M_s$ is the saturation magnetization, $\lambda$ is the saturation magnetostriction of the material and $\sigma$ is the stress. Strictly speaking, this relation would hold if the magnetization within the nanomagnet is spatially invariant and equal to the saturation magnetization everywhere, which would be the case for a single domain nanomagnet, but not for a nanomagnet with multiple (complex) domains as these nanomagnets are. Nonetheless, this is a convenient quantity for calibration. For Terfenol-D nanomagnets, the effective field for the maximum applied stress of 50 MPa (calculated in this fashion) is 560 Oe. For Terfenol-D nanomagnets, the effective field for the maximum applied stress of 50 MPa is 560 Oe.

In Fig. 7, we plot the switching error probability as a function of applied stress amplitude for *cobalt* nanomagnets, where once again the stress was applied for a fixed duration of 1.5 ns. The maximum stress amplitude that we consider for cobalt nanomagnets is 250 MPa. The Young's modulus of cobalt is 209 GPa and hence a stress of 250 MPa will produce a strain of 1200 ppm, which is about the maximum a cobalt nanomagnet can sustain before strain is relaxed through the formation of dislocations. For Co nanomagnets, the effective field $H_{eff}$ for the maximum applied stress of 250 MPa is 120 Oe.

We do not plot the results for defects C2 and C3 in the case of cobalt because the shape anisotropy energy barrier in cobalt nanomagnets with these defects becomes so low that just the bias magnetic field can always switch the magnetization at room temperature *without* the need for any stress to depress the shape anisotropy energy barrier. This is an interesting observation; the defect-free nanomagnet will not switch without stress, but the defective nanomagnets with C2 and C3 types of defects do because the energy barriers in these nanomagnets are lower than that in a defect-free nanomagnet and low enough for the 3 mT magnetic field to overcome. This is one example where the defects may have a beneficial effect depending on the application.

In every plot in both Figures 6 and 7, there is a range of stress where the switching error probability is *minimum*. This range is obviously where the "critical stress" is. Both under-stressing and over-stressing result in higher error probability than what is observed in the critical stress regime. Thus, this observation is consistent with the discussion in Section V.

Note also that "localized" defects (e.g. a hole or a bump in the center – C1, C5 or C6) are less harmful than "delocalized/extended" defects (different thicknesses in different halves, or a rim – C2, C3, C4), because the error probabilities in the latter cases are much higher. The rim defect (C4) seems to be the worst offender producing the highest error probability.

In Figure 8, we plot the switching error probability as a function of the acoustic wave stress amplitude for each defect in cobalt, except C2 and C3, and for two different stress wave frequencies – 200 MHz and 1 GHz. We do not do this for C2 and C3 since these nanomagnets have very low energy barriers and their magnetizations can be flipped by the magnetic field alone, without the need for any stress. We also examine only two frequencies (and not more) because of the excessive computational burden. For example, to simulate for one period at 200 MHz, we have to simulate for 5 ns with a time step of 0.1 ps. This requires 50,000 time steps. We have to repeat this for 1000 switching trajectories, which requires $50,000 \times 1,000 = 5 \times 10^7$ steps. Then, we have to repeat the entire exercise for 5 different nanomagnets with four different stress amplitudes, requiring $20 \times 5 \times 10^7 = 10^9$ steps for each frequency. This takes several weeks of run time on a special purpose computer for running MuMax3. Therefore, examining too many frequencies will be computationally prohibitive.

It is intriguing to find from Fig. 8 that for the lower frequency, time varying stress (of sufficient amplitude) is actually *more effective* in switching nanomagnets (both defect-free and defective) than static stress since the error probabilities are lower for all cases, except defect C4. For the 200 MHz sinusoidal stress, the error probability is < 0.001 (we simulated only 1000 trajectories and none of them failed) for all defects except C4, as long as the stress amplitude exceeds a threshold value. This is a promising result since it shows that switching with sinusoidal stress results in low error probabilities even in the presence of most types of defects. What this reveals is that gradual stressing and de-stressing is more effective than abrupt stressing and de-stressing, provided the frequency is slow enough (200 MHz) to allow the magnetization enough time to respond to the stress. If the frequency is too high (1 GHz), the magnetization does not have ample time to respond to the stress and this increases the failure probability. The defect C4 is an exception where, for any given stress amplitude, the slow frequency results in *higher* error probability than the fast frequency. This tells us that the interplay between the temporal dynamics of magnetization and the temporal dynamics of stress is complex and that could result in counter-intuitive results in some cases.

It is interesting to observe that in the case of high frequency (1 GHz), the switching error shows a non-monotonic dependence on stress amplitude and is higher at 200 MPa than at all other stress amplitudes. The exception to this is C4 for which the switching error is smaller at 200 MPa than at all other stress amplitudes considered. This, and other anomalous behaviors, sets C4 apart from other defects. The defect C4 is actually a frequent occurrence and it always results in a relatively high error probability.

What makes C4 particularly debilitating can only be speculated on; it is very likely the nature of the potential energy profile within this nanomagnet. The potential profile (which would vary spatially) may be too high in some regions for any reasonable amount of stress to overcome, leading to large error probability. It is also possible that this defect spawns metastable states (local energy minima in the potential profile) that pin the magnetization. We cannot compute the potential energy profile, but in Fig. 9, we show the micromagnetic distributions in nanomagnets C4 and C6 at equilibrium (when the only fields present are the shape anisotropy field and the exchange interaction field). We see edge domain formation in C4 (not seen in C6 which appears virtually single-domain), which extend outward from underneath the rim, and that may contribute to C4 being more error-prone.

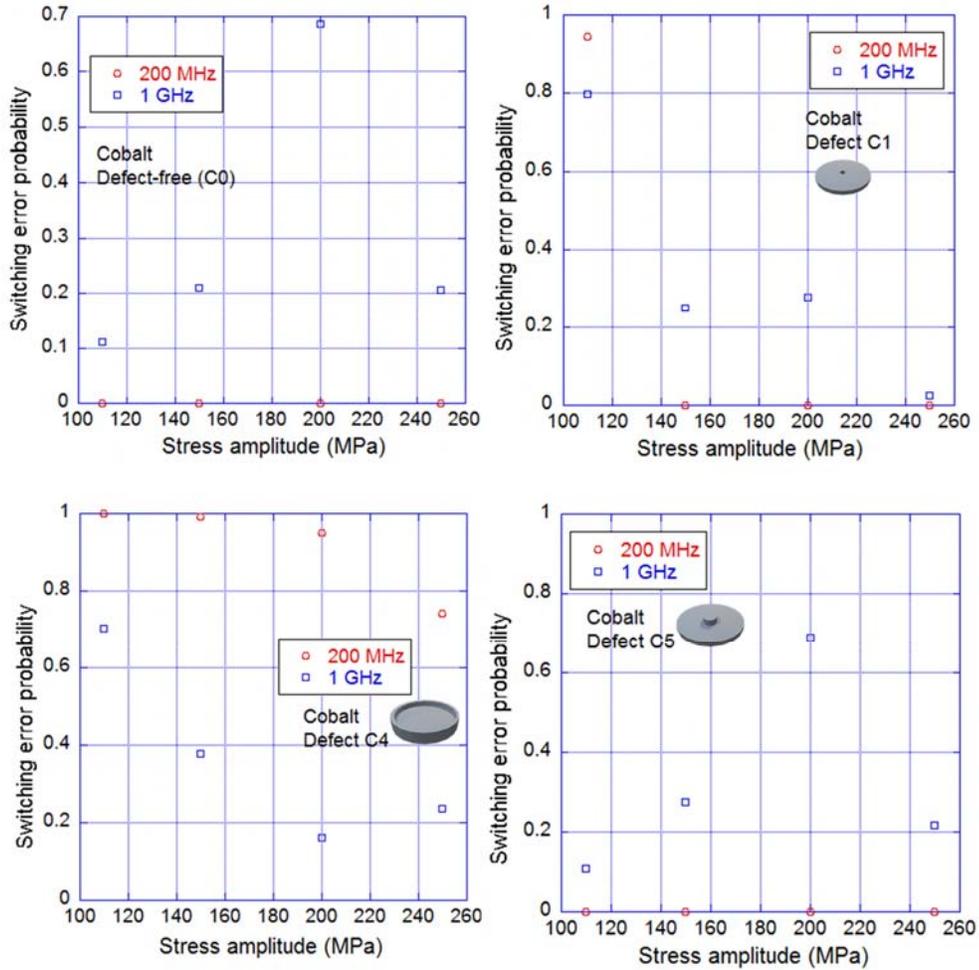

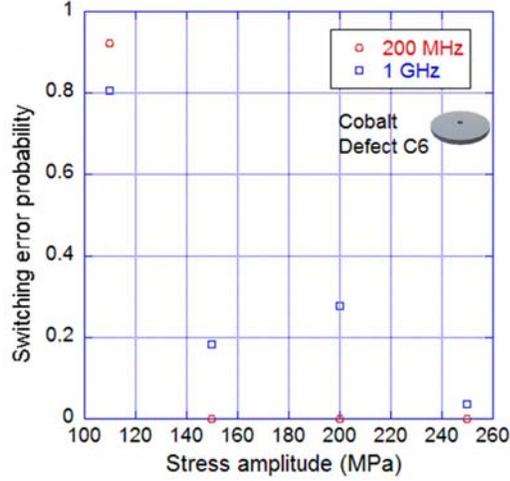

Fig. 8: Switching error probability at room temperature as a function of applied uniaxial stress amplitude of a surface acoustic wave of frequency 200 MHz and 1 GHz for a cobalt elliptical nanomagnet of nominal major axis 100 nm, minor axis 90 nm and thickness 6 nm. The plots are for the defect-free and defective nanomagnets with defects C1, C4, C5 and C6.

The non-monotonic dependence of the error probability on stress amplitude for all nanomagnets except C4 is very likely a manifestation of the interaction between stress amplitude and stress frequency. We would normally expect the error probability to increase with stress amplitude beyond critical stress, as discussed earlier, but the reason why the error probability drops at the highest stress considered (250 MPa) has a more complex origin. When the nanomagnet is stressed, there is a distribution of the magnetization orientation $f(\theta)$ centered at $\theta = 90^0$ where the potential well is (see the middle panel of Fig. 5). When stress is released and the potential barrier is restored at $\theta = 90^0$, the segment of the distribution that is to the right of $\theta = 90^0$, i.e. in the interval $90^0 \leq \theta \leq 180^0$, will contribute to switching success, whereas the segment that is to the left of $\theta = 90^0$, i.e. in the interval $0^0 \leq \theta \leq 90^0$ will contribute to switching failure. This was discussed in refs. [16, 34] and will not be repeated here since it is outside the scope of this work. This distribution depends on the interplay between the stress amplitude and stress frequency and is least favorable at 200 MPa stress amplitude, which is why the error happens to be maximum at that amplitude.

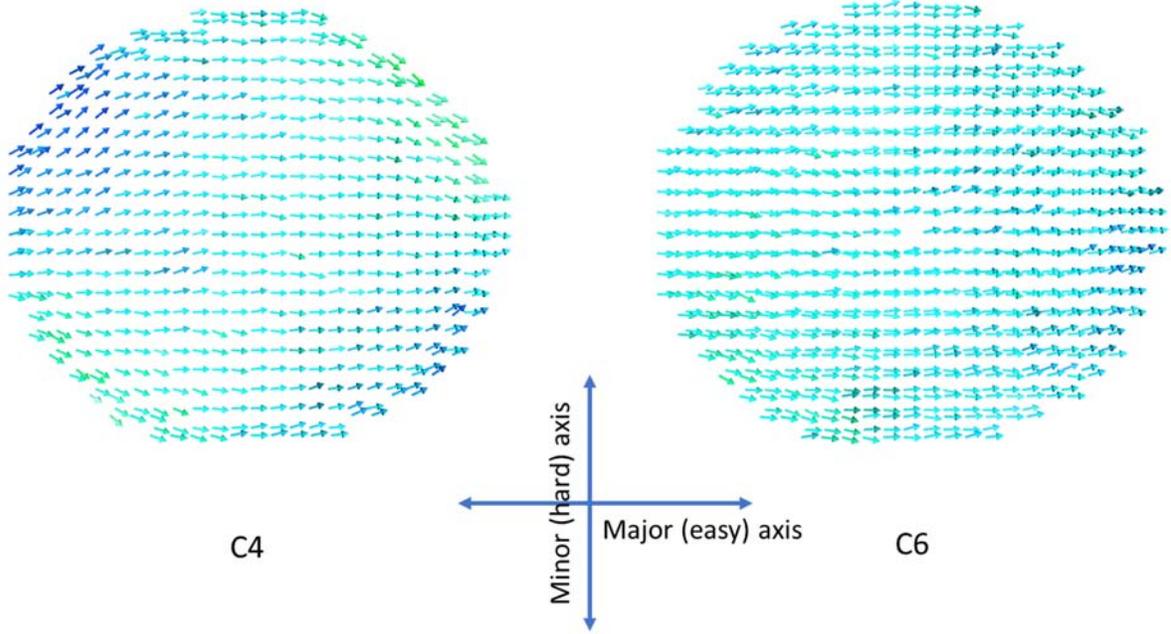

Fig. 9: Micromagnetic distributions within nanomagnets of type C4 and C6 at equilibrium.

Finally, we also obtained the magnetization curves (M vs. H) using MuMax3 for the seven different nanomagnets to observe how defects affect this characteristic. To obtain the M-H plots, we first allow the micromagnetic distributions to reach equilibrium and then gradually ramp up the magnetic field in steps of 10 Oe to generate the M-H plot. Some nanomagnets already have saturation magnetization in the absence of an external field (H = 0). The plots are shown below for both Terfenol-D and Cobalt in Fig. 10.

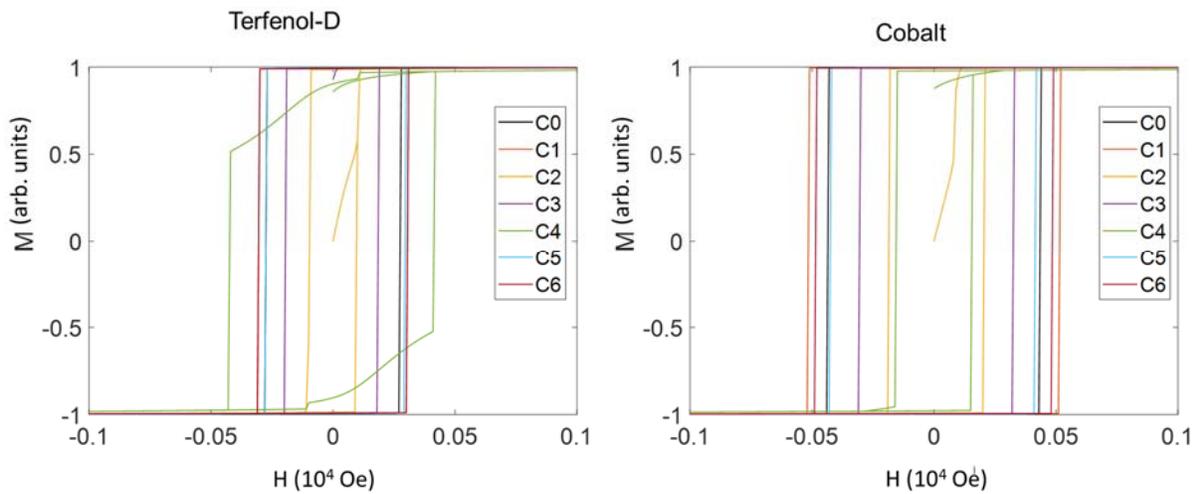

Fig. 10: Simulated magnetization curves (M- vs. H) for the defective and defect-free nanomagnets.

Fig. 10 shows that defects can either increase or decrease the coercivity of the nanomagnets and therefore affect the switching characteristics if we were switching with a magnetic field instead of magneto-elastic coupling. Thus, defects impact not just magneto-elastic switching, but perhaps other switching modalities as well, although a study of that is outside the scope of this work. Defect C4 once again stands out from the rest in the case of Terfenol-D since unlike the others, it does not exhibit a rectangular hysteresis curve.

**Conclusion**

The results reported in this paper show that structural defects, usually acquired during the fabrication process, can significantly increase the switching error probabilities of nanomagnets that are switched magneto-elastically, sometimes by orders of magnitude. They can also vastly increase the critical stress values that produce the lowest error probability. While defect-free nanomagnets show very low error probability in the critical stress regime (<0.001), defective nanomagnets show error probabilities that are much larger. This could explain the differences in error probabilities calculated theoretically (assuming defect-free nanomagnets) and observed experimentally, since real nanomagnets used in experiments very likely have some of these defects.

More importantly, the switching error probabilities in defective nanomagnets are so large that they exceed what would be acceptable for Boolean logic *by several orders of magnitude*. *This calls into question the very viability of magneto-elastic (or straintronic) Boolean logic*. The error rates may be too high for even memory. This is the main conclusion of this work. There is considerable enthusiasm surrounding magneto-elastic (straintronic) logic because it is believed to be extremely energy-efficient and therefore offers a potential solution to extending Moore's law, but the results of this paper should temper some of that enthusiasm.

The magneto-elastic community often believes that employing materials with higher magnetostriction (e.g. Terfenol-D instead of cobalt) will lead to a larger effective switching field due to stress anisotropy and that may improve the switching probability. This may not be necessarily true when one considers bi-directional coupling between the piezoelectric and magnetostrictive components [51], but even if that were true, defects can *erase* any advantage of the higher magnetostriction material. For example, when defect of the type C4 is present, the switching error probability in the Terfenol-D nanomagnet turned out to be actually worse than that in the cobalt nanomagnet, even though Terfenol-D has much higher magnetostriction. This shows that it is more imperative to eliminate defects than to develop better magnetostrictive material because defects have a much larger influence on switching error rates than the material (magnetostriction).

We also found that *localized* defects (such as a small hole or hillock) are more forgiving than *delocalized* (extended) defects such as thickness variation across a significant fraction of the nanomagnet surface, or rim around the periphery. However, even localized defects give rise to switching error probabilities that are too high for error-intolerant computing applications such as Boolean logic and memory. This seems to suggest that magneto-elastic switches may be better adapted to certain types of non-Boolean computing paradigms, particularly collective computational models where the collective activity of many nanomagnets working collaboratively elicit the computational activity (e.g. image processing [27, 28]). There, the failure of a single (or few) switch is not catastrophic.

Finally, we found that sinusoidal time varying stress of the right frequency can reduce the switching error probability compared to constant stress for most cases. This is consistent with the observation in ref. [44] which found a very high success rate with time-varying stress. Thus, if magneto-elastic switches are to be used in memory, writing with a time-varying stress [39-50], as opposed to a stress pulse of fixed amplitude, may be beneficial.

**Acknowledgement**: This work was supported by the US National Science Foundation under grants ECCS-1609303 and CCF-1815033. D.W. acknowledges the Dean's Early Research Initiative program for the College of Engineering at Virginia Commonwealth University. S.S. acknowledges the S. N. Bose National Center for Basic Sciences for support.